\pdfoutput=1
\documentclass[preprint,aps,prb]{revtex4}
\usepackage[dvips]{graphicx}
\usepackage{color,array,dcolumn}
\usepackage{amsmath}
\usepackage{amssymb}

\begin{document}

\title{Hydrogen bonding characterization in water and small molecules}

\author{Pier Luigi Silvestrelli}
\affiliation{Dipartimento di Fisica e Astronomia, 
Universit\`a di Padova, via Marzolo 8, I--35131, Padova, Italy}

\begin{abstract}
\date{\today}
The prototypical Hydrogen bond in water dimer and Hydrogen bonds 
in the protonated 
water dimer, in other small molecules, in water cyclic clusters, and in ice,
covering a wide range of bond strengths, are
theoretically investigated by first-principles calculations based on the
Density Functional Theory, considering a standard Generalized Gradient 
Approximation functional but also,
for the water dimer, hybrid and van-der-Waals corrected functionals. 
We compute structural, energetic, and electrostatic (induced molecular
dipole moments) properties.
In particular, Hydrogen bonds are characterized in terms of differential 
electron densities distributions and profiles, 
and of the shifts of the centres of Maximally localized Wannier 
Functions. The information from the
latter quantities can be conveyed into a single geometric bonding 
parameter that appears to be correlated to the Mayer 
bond order parameter and can be taken as an estimate of the 
covalent contribution to the Hydrogen bond. 
By considering the cyclic water hexamer and the hexagonal
phase of ice we also elucidate the importance of cooperative/anticooperative
effects in Hydrogen-bonding formation. 
\end{abstract}

\maketitle

\section{Introduction}
The Hydrogen bond (HB), namely the pairwise interaction between
an electronegative atom and a H covalently bound to another
electronegative atom, is a relatively weak (often of the order of
100-300 meV/HB\cite{Jeffrey,Walrafen}) bond of 
paramount importance, being the most effective of all directional 
intermolecular interactions. For instance, it is the
interaction which determines the peculiar properties
of water in the condensed phases, and holds together the
two strands of DNA in the double helix and the 3-dimensional
structure of proteins.\cite{Jeffrey}
Moreover, HBs play a vital role in the chemistry of life because they
can easily be made and broken at ambient temperatures.\cite{Buckingham}
Actually there are several different types of
HBs and dissociation energies span more than 2 orders of magnitude,
from 0.2 to 40 kcal/mol (from 0.01 to 1.7 eV); this wide range means
that different types of interaction energies, i.e. electrostatic, 
induction, electron delocalization, exchange repulsion and dispersion 
give different relative contributions to the overall energy of the HB;
however, for the vast majority of normal HB in the vapor and
condensed phases, the electrostatic-plus-induction description
seems to provide a near quantitative account of the attractive force
responsible for the interaction.\cite{Buckingham,Steiner}

The directional interaction between water molecules is the
{\it prototype} of all HBs and can be termed the ``classical HB'' 
(of moderate strength):\cite{Steiner}
the O-H bonds of a water molecule are inherently polar
with partial atomic charges of around +0.4$e$ on each H atom and -0.8$e$ 
on O. In water
the intermolecular distance is shortened by around 1 \AA\ compared to the
sum of the van der Waals (vdW) radii for the H and O atoms, which indicates 
a substantial overlap of electron orbitals to form a 3-center, 4-electron 
bond; despite significant charge transfer in the HB of the water dimer, 
the total interaction is assumed to be dominantly electrostatic.\cite{Steiner}
The directionality of moderate and weak HBs is relatively soft but
can still be identified with the orientation of electron lone pairs.
A very important way\cite{Steiner} of looking at HBs is to regard them as
{\it incipient proton transfer reactions}, so that, in a stable HB,
X-H...Y a partial bond H...Y is already established and the X-H bond is
concomitantly weakened. 
In normal HBs the covalent X-H bond lengthens slightly but
remains intact in the complex; if further proton-transfer occurs,
the resulting complex becomes a H-bonded ion pair, the Y-H 
distance approaches the Y-H distance in the monomer cation, the
X-H distance becomes much longer than the covalent X-H distance
and the X-Y distance lengthens slightly.\cite{Buckingham}  
Interestingly, in a chain with 2 HBs both become stronger.
This effect, that is further amplified by considering chains
of multiple HBs (as in the cyclic form of the hexamer water cluster)
or an extended HB network (as in condensed phases of water, such as 
hexagonal ice), can be ascribed to $\sigma$-bond {\it cooperativity}
since the charges flow through the X-H $\sigma$ bonds leading to
``polarization-enhanced H bonding'',\cite{Steiner} and increased
intermolecular orbital interactions.\cite{BWang}

A long-standing controversy exists about the theoretical origins of 
H-bonding, whether 
primarily due to classical electrostatics (ionic or dipole-dipole forces) 
or quantum covalency (see, for instance refs. 
\onlinecite{Barbiellini,Romero,Weinhold,Chaplin,Sterpone} 
and references therein).
If the water HB is considered within the context of the complete 
range of molecular H bonding then it appears most probable that it is not
solely electrostatic.\cite{Chaplin} 
A proton NMR resonance experiment\cite{Elgabarty} confirmed the covalency of
HBs in liquid water and theoretical calculations\cite{BWang} showed
that delocalized molecular orbitals exist in water rings.\cite{BWangSR}
Clearly, the interaction between two nuclei can be considered a true
chemical bond only if there is a concentration 
of electron density around the internuclear axis. 
Results from recent theoretical and experimental investigations
(in particular with Atomic Force Microscopy) 
suggest that the HB has both an electrostatic origin and a 
partly covalent character.\cite{Zhang} Moreover,
extensive investigations based on the ``natural bond order'' (NBO) 
method\cite{Weinhold,NBO} led to conclude that, although in HBs
classical electrostatic effects are undoubtedly present, they
seem to play only a secondary role with respect to non-classical 
resonance effects represented by
intermolecular bond order or charge transfer.\cite{Weinhold}
However, very recent studies cast doubts on the validity of the NBO
method to evaluate charge-transfer effects in intermolecular interactions
and hence on their predominance in characterizing the HBs.\cite{Stone} 
In any case their role in H bonding is probably not less significant 
than those of polarization and frozen density interactions.\cite{Khaliullin}

The prototypical HB in water dimer and HBs in the protonated 
water dimer, in other small molecules, in water cyclic clusters, and in ice,
covering a wide range of bond strengths, are here
theoretically investigated by first-principles calculations based on the
Density Functional Theory (DFT).
We compute structural, energetic, and electrostatic (induced molecular 
dipole moments) properties, also testing, in the water dimer case,
hybrid and van-der-Waals corrected functionals.
In particular, HBs are characterized in terms of differential 
electron density distributions and profiles, and of a geometric bonding
parameter defined in terms of the positions of the centres of Maximally 
localized Wannier Functions.  
This parameter appears to be correlated to the Mayer 
bond order parameter and the HB strength, and gives an estimate of the 
HB covalent character. 
By considering the cyclic water hexamer and the hexagonal
phase of ice we also elucidate the importance of cooperative/anticooperative
effects in H-bonding formation. 

\section{Method and Computational details}
We perform first-principles calculations within the framework of the DFT.
DFT represents the most popular theoretical method to
investigate the structural and electronic properties of molecules and
condensed matter systems from first principles, however the
ability of DFT to quantitatively describe HBs between water molecules
in either small clusters or the liquid state is not definitively
assessed yet,\cite{Santra1,Santra2} although some
Generalized Gradient Approximation
(GGA) functionals, such as BLYP\cite{BLYP} and PBE,\cite{PBE} are
commonly (and often successfully) used in
DFT-based simulations of water, since they seem to offer a good compromise
between computational efficiency and accuracy.
In principle, long-range electron correlation interactions, attributed
to van der Waals (vdW) forces, can contribute significantly to HB binding
energies;\cite{mycpl,Arey} however, vdW interactions are not properly
described\cite{Kohn} by current GGA density functionals.
Hence, in the conventional GGA descriptions of
HBs, presumably electrostatic, exchange, and
induction effects somehow compensate for the spurious or insufficient
accounting of vdW long-range correlation forces.

Obviously, there is a strict relation between the electron density 
topology and physical-chemical properties and this can also be made 
more precise and quantitative by referring to the well-known DFT
Hohenberg-Kohn theorem,\cite{HK} which asserts that the ground-state properties
of a system are a consequence of its electron density.
Therefore, we expect that, upon bonding, similar electron-charge 
redistributions basically correspond to the same type of
interaction,\cite{Lane} so that a detailed charge analysis represents
a very useful tool for intermolecular-bonding characterization.
To evaluate the electron density response due to the bond formation between
different fragments we compute the {\it differential} charge density,
$\Delta \rho$,
defined as the difference between the total electron density of the system
and the superposition of the densities of the separated fragments 
(atoms or molecules), keeping the 
same geometrical structure and atomic positions that these
fragments have within the optimized system.
This is a meaningful procedure since, for instance, the geometrical
distortion of the two water molecules in a water dimer is
insignificant.\cite{Khaliullin}
By drawing $\Delta \rho$ isosurfaces on a planar plot we highlight
the electron charge redistributions related to the HB formation. 
Moreover, one-dimensional profiles $\Delta \rho (z)$ can also 
be obtained, which are very effective for describing charge modifications 
upon H bonding: $\Delta \rho (z)$ is computed 
along an intermolecular $z$ axis, as a function of
$z$ values, by integrating $\Delta \rho$ over the corresponding, 
orthogonal $x,y$ planes.

We also rely on the use of the Maximally-Localized
Wannier function (MLWF) formalism,\cite{Marzari} that allows
the total electronic density to be partitioned, in a chemically
transparent and unambiguous way, into individual fragment contributions.
The MLWFs, $\{w_n({\bf r})\}$, are generated by performing a unitary
transformation in the subspace of the occupied Kohn-Sham orbitals,
obtained by a standard DFT calculation, so as to minimize the total spread:
                                                                                
\begin{equation}
\Omega = \sum_n S_n^2 =
\sum_n \left( \left<w_n|r^2|w_n\right> -
\left<w_n|{\bf r}|w_n\right>^2 \right)\;.
\label{spread}
\end{equation}
Besides its spread, $S_n$, each MLWF is characterized also by its
Wannier-function center (WFC), defined as the center of mass of the
(square modulus) Wannier function.
Note that, if spin degeneracy is exploited, every MLWF corresponds
to 2 paired electrons.
Knowledge of WFC positions also allows to estimate molecular dipole
moments,\cite{psil1999} and the MLWFs can be even used to include 
vdW interactions in DFT.\cite{mycpl,Silvestrelli}

Calculations have been performed mostly
with the Quantum-ESPRESSO ab initio package\cite{ESPRESSO} and the
MLWFs have been generated as a post-processing calculation using
the WanT package.\cite{WanT}
For our calculations on molecules and clusters we have adopted a 
periodically-repeated, simple cubic supercell, with a side of 16 \AA, 
in such a way to make the interactions among periodic
replicas negligible.
Electron-ion interactions were described using ultrasoft
pseudopotentials and the sampling of the Brillouin Zone was limited to the 
$\Gamma$-point.
Instead for ice in the hexagonal phase we used an hexagonal supercell 
containing 12 water molecules at experimental density
and with a $2\times2\times2$ $k$-point sampling of the Brillouin Zone.

\section{Results and Discussion}

\subsection{Water dimer}
As can be seen looking at Fig. 1 and 2, showing the differential 
electron charge density, $\Delta \rho$, 
plotted on a 2D plane and integrated over $x,y$ planes orthogonal to 
the intermolecular O-O $z$ axis ($\Delta \rho (z)$ function), respectively, 
there is no overall electron-charge accumulation between the two O atoms
of the water dimer in the optimal, linear HB configuration.
Actually, one can only observe (see notation defined in Fig. 3) 
a small amount of electron-charge {\it redistribution}: 
an electron depletion of about 0.008 $e$ (that is less than 
1\% of an electron) between the O atom of the {\it acceptor}
molecule, O$_a$, and H (in the HB region) is exactly offset by
an electron accumulation between H and the O atom of the {\it donor} 
molecule, O$_d$, while a more substantial (about 0.03 $e$) charge transfer 
appears to occur from O$_a$ to O$_d$. This is compatible with previous 
findings.\cite{Weinhold,Khaliullin,Galvez,Lee,Jiang,Nilsson}
The fact that the acceptor molecule contributes more than the donor is 
in line with the Lewis concept of acid/base behavior.\cite{Galvez} 
As previously observed,\cite{Nilsson} the amount of charge accumulated
in the middle of the HB is 2 orders of magnitude smaller than in 
a typical covalent bond, such as, for instance, in the H$_2$ dimer, thus
showing that in the water-dimer HB the electron pairing contribution
is probably marginal and suggesting that instead 
charge polarization of one of the lone-pair
orbital of the acceptor molecule towards the H atom of the donor molecule
plays a key role.\cite{Nilsson}
Note also the pronounced electron-density depletion at the position
of the H atom participating in the HB which therefore increases
its positive charge.

Interestingly, Hartree-Fock (HF) methods do not give\cite{Nilsson} 
strong evidence of charge transfer involving the lone pair orbital 
in contrast to results from DFT methods, in line with the substantial
HF underbinding of the water dimer; instead DFT in different GGA
flavors (such as PBE adopted here) accurately reproduces the water-dimer 
binding energy and bond length.   
In fact our PBE results for basic energetic and structural parameters of 
the optimal configuration of the water dimer
(binding energy = -221 meV, O$_a$-H distance = 1.92 \AA, 
H-O$_d$ distance = 0.98 \AA, O$_a$-H-O$_d$ angle = 178$^o$, H being
the H atom involved in the HB) are in excellent agreement with literature
reference data\cite{Tschumper}   
(binding energy = -218 meV, O$_a$-H distance = 1.95 \AA, 
H-O$_d$ distance = 0.97 \AA, O$_a$-H-O$_d$ angle = 173$^o$).
From these considerations we expect that our analysis on 
redistributions of the electronic charge, at the PBE level, is meaningful 
for a proper description of realistic H-bonded systems (particularly for
water). 
  
In Fig. 4 we report plots of $\Delta \rho (z)$
for the water dimer, obtained by DFT functionals
different from PBE: vdW-DF-cx,\cite{BerlandPRB,BerlandJCP} which is one
of the most recent versions of the vdW-DF family where
vdW effects (not properly described by GGA functionals) are included by
introducing DFT nonlocal correlation functionals, and PBE0\cite{PBE0} and
B3LYP,\cite{B3LYP} which are {\it hybrid} functionals where a given 
fraction of exact exchange is combined with the PBE or BLYP GGA functional,
respectively: PBE0 is particularly accurate in describing H-bonding
in water clusters,\cite{BWang} while
B3LYP provides a better description of exchange in the 
intermolecular region than DFT approaches based on the GGA.\cite{Fuster}
As can be seen, differential density profiles generated by these alternative
functionals are very close to that obtained by PBE, so that also
the amounts of the electronic charge redistributions are quite similar,
thus showing that, for this system, vdW effects or inclusion or a
fraction of exact exchange are not essential and PBE performs quite well,
thus further confirming the previous considerations.
The small impact of vdW interactions in moderate intermolecular
HBs have been also recently reported.\cite{Katsyuba}

As a result of the electron-charge redistribution and polarization effects
described above, the dipole moments
of the acceptor and donor water molecules (estimated following ref.
\onlinecite{psil1999}) are changed with respect to that of the single
water molecule ({\it induced} dipole moments); 
moreover the dipole moment is slightly 
larger (2.18 D) for the acceptor molecule than for the donor one (2.00 D), 
the dipole moment of the isolated water molecule being 1.86 D. 
In particular, the larger dipole moment of the acceptor molecule originates
from the small charge extrusion from O$_a$ towards the H atom of the 
donor molecule (see Fig. 1). 
In summary, for the water dimer the nature of the HB can be 
characterized in terms of {\it quantum} effects associated to 
{\it redistribution} rather than transfer of electronic charge, in line
with the conclusions of previous studies (see, for instance ref.
\onlinecite{Galvez}): in fact a quantum description is required to
describe the small deformations of the electron distributions of the
water monomers when they interact, however the scenario is clearly
different from that of a typical covalent bonds.
In view of the above considerations, the success of semiempirical
(often non-polarizable) models in reproducing many of the properties
of water, that are obviously unable to properly describe charge-transfer
and quantum-mechanical effects, appears to be to some extent 
fortuitous.\cite{Lee} 
A partial explanation for the good performances of these models, 
despite the importance of charge transfer for the dimer, could be
ascribed\cite{Lee} to the fact that charge transfer occurs for the H-bonded
water dimer because there is an evident asymmetry between the two 
molecules as one 
donates and one accepts a HB, while in the condensed phases each 
molecule is in a more symmetric local environment.\cite{Lee} However,
if symmetry is broken, for instance by the addition of a solute or the creation 
of an interface, then charge transfer would probably become more 
important and semiempirical models less reliable.\cite{Lee} 
For the sake of completeness,
we should point out that water models exist (see, for instance, ref.
\onlinecite{Jiang}), where the effects of charge transfer are included
by adding an explicit H-bonding term. 

In order to further characterize H-bonding in water dimer in terms 
of differential electron charge density, we have also studied
what happens if the water-water distance is reduced 
(O-O distance = 2.5 \AA) or increased (O-O distance = 4.0 \AA) with
respect to the equilibrium value (O-O distance = 2.91 \AA), see
curves in Fig. 5.                        
One expects that decreasing the O-O distance would make the HB more 
covalent in character,\cite{Galvez} while the covalent contribution 
and charge-transfer effects should be negligible at distances significantly 
larger than equilibrium.\cite{Lee}
As can be seen, between O$_a$ and the H atom involved in the H-bonding 
(that is in the HB region), at short water-water distance there is
a more pronounced $\Delta \rho (z)$ depletion
than at equilibrium distance, which does not support a H-bonding description
in terms of covalent character but instead mainly a polarization mechanism,
that is a process of separating opposite charges within the system; at the
same time, more significant density 
accumulation takes place between H and O$_d$ (that is in the covalent bond 
region) and in the region above O$_d$, while more pronounced density depletion
is observed below O$_a$.
Instead, at long water-water distance, in all the regions $\Delta \rho (z)$
is always quite reduced (note that,
also in this case vdW effects do not seem to play a key role since 
replacing the PBE functional with vdW-DF-cx does not lead to significant
changes).  

\subsection{Water hexamer, ice, and cooperativity}
A special feature of the HB is its {\it cooperativity}, i.e., the fact 
that the local HB strength is influenced by the neighboring water 
molecules as a consequence of 3-body effects.\cite{Perez} 
In particular, a water molecule with 2 HBs whereit acts as both donor
and acceptor is somewhat stabilized relative to one where it is either the
donor or acceptor of 2 HBs,\cite{Chaplin} so that, for instance,
the formation of one HB in a H-bonded chain
cooperatively enhances the strength of another HB.
Both proton donor (D) and acceptor (A) participating in a H-bonding
pair DA are capable of forming H bonding with the other water
molecules; D can additionally accept two protons and donate one proton, and
A can additionally donate two protons and accept one proton.
A classification of H-bonding patterns considering the cooperativity
has been proposed,\cite{Ohno} based on the parameter $d'a'DAd"a"$, 
where $d$ and $a$ are integers
indicating the number of proton donors and acceptors to $D$ (the {\it single
prime}) and $A$ (the {\it double prime}), respectively. Then, a magnitude given
by MOH =$ -d' + a' + d" -a"$  has been introduced, assuming that each 
contribution for the enhanced ability of the HBs is the same and 
each effect is additive. 
The possible values for MOH are therefore: -2, -1, 0, 1, 2, 3, and 4.
Note that HBs may not only enhance but also reduce the strengths of 
each other, which is probably responsible for the preferences of 
{\it homodromic} over {\it antidromic} cycles of HBs: in the preferred 
{\it homodromic} arrangement all HBs run in the same direction, while in 
the less common {\it antidromic} arrangement a change of orientation leads to 
local {\it anticooperativity}.\cite{Steiner,Saenger}

In order to investigate cooperative and anticooperative effects of
H-bonding in water, we have studied two different conformations 
of the {\it cyclic} water hexamer (see Figs. 6 and 7). The first one is
the usual {\it homodromic} configuration where all the HBs run in the 
same direction, leading to strong cooperative effects 
(MOH = 2 for all the HBs, which gives a total sum of 12), 
while, in the other {\it antidromic} one, some water
molecules are rotated giving rise to an anticooperative effect 
(MOH = 2 for only 2 HBs, while MOH = 0 for the others, which gives a 
total sum of 4). 
The corresponding, integrated differential densities $\Delta \rho (z)$
are plotted in Fig. 8 where also the water-dimer profile is reported. 
For a proper comparison with the single HB of the dimer case of Fig. 2,
$\Delta \rho (z)$ 
for the hexamers is obtained by choosing as the $z$ axis the vertical
axis of a {\it cylinder} centered around a selected O-O segment (connecting 2
adjacent water molecules) and with a radius of 2.3 \AA\ (to essentially
make differential-density contributions from other pairs of water molecules 
negligible); $\Delta \rho (z)$ is then obtained by integrating over
planar regions orthogonal to $z$ and contained inside the cylinder.  

As can be seen, in the two water-hexamer cases the $\Delta \rho$
distribution looks quite different: in Fig. 6, for a given pair of 
water molecules, it resembles that of the water dimer shown in Fig. 1, while 
in Fig. 7 it becomes less symmetric (due to 
cooperative/anticooperative effects on different water molecule) and there is
an appreciable accumulation in the HB region only for those 
HBs that are characterized by nonzero MOH values (MOH = 2) and are
located in the top-left and bottom-right regions of the hexagon. 
These observation are confirmed and made more quantitative by the 
$\Delta \rho (z)$ curves plotted in Fig. 8. 
In particular, the curve corresponding to the {\it homodromic} hexamer case is 
similar to that of the water dimer (see Fig. 2 and Fig. 5, particularly
for the O-O distance shorter than at equilibrium), however, as expected, the 
cooperative effects tend to amplify the $\Delta \rho (z)$ fluctuations
and charge redistributions; clearly in this configuration the distinction 
between $O_d$ and $O_a$ disappears since all the O atoms are simultaneously 
HB donors and acceptors. Instead, in the {\it antidromic} hexamer case,
anticooperative effects tend to suppress $\Delta \rho (z)$ fluctuations
and reduce charge redistributions (here the 
$\Delta \rho (z)$ profile is relative to the pair of water molecules on the 
left side of the hexagon shown in Fig. 7 and the bottom O atom is only
acceptor while the top O atom is both donor and acceptor).

The ``polarization-enhanced H bonding''\cite{Steiner} due to cooperative 
effects can be quantitatively evaluated by computing the average HB
energy $E_b$ (defined as the water dimer or hexamer binding energy 
divided by the number of HBs, 1 or 6, respectively) and significantly 
affects also interatomic distances and dipole moments.  
In fact,  $E_b$ = -221, -381, -228 for the dimer, the
{\it homodromic} hexamer and the {\it antidromic} hexamer, respectively. 
This means that cooperative effects in the favored cyclic hexamer 
configuration lead to an energy gain of more than 70\% (note that 
similar effects are also observed in the ring structure of 
methanol hexamer\cite{Kashtanov}). 
Instead, in the {\it antidromic} case
cooperative effects are essentially suppressed and the absolute value of
$E_b$ is only marginally increased with respect to the water dimer case. 
As it is well known (see, for instance, ref. \onlinecite{Perez})
cooperative effects are also revealed by the nearest neighbor O-O distances
of H-bonded water molecules, whose average values are
2.90, 2.66, and 2.97 \AA\ for the dimer, the
{\it homodromic} hexamer and the {\it antidromic} hexamer, respectively 
(the corresponding HB O...H distances are $d_b = 1.92$, 1.62, 1.92 \AA). 
Hence the O-O distance decrease with respect to the dimer provides direct 
evidence for cooperativity in {\it homodromic} hexamer,\cite{Buckingham} and
confirms that, for a given substance such as water, 
shorter O-O (and O...H) distances generally correspond to stronger HBs.

HB cooperativity also leads to a significant enhancement of the
water-molecule dipole moments as a consequence of polarization effects; 
in fact, the average dipole-moment values are 2.09, 3.05, 2.31 D 
for the dimer, the {\it homodromic} hexamer and the {\it antidromic} hexamer, 
respectively (we remember that
the dipole moment of the isolated water molecule is 1.86 D). Hence,
in the {\it homodromic} hexamer there is a dipole increase 
of almost 50\% with respect to the dimer case (and of more than 60\% 
with respect to the isolated water molecule), while in the {\it antidromic} 
case the dipole increase is considerably reduced (about 10\%). 
These findings on cyclic water structures are important also because
the exploration of structural and bonding properties of small water
clusters provides a key for understanding anomalous properties of
water in the liquid and solid phases.\cite{Ohno}

We have also investigated the case of ice in the standard {\it hexagonal}
phase, where, in the ideal structure, all the HBs are characterized by
MOH = 2, thus suggesting again substantial cooperative effects.
As can be seen in Fig. 8 (two-dimensional plots for HBs in 
ice are not reported since 
they closely resemble those relative to water dimer and cyclic hexamer),
the $\Delta \rho (z)$ profile turns out to be intermediate between that
of water dimer and that of {\it homodromic} hexamer and the same is true for the
HB energy: $E_b=-328$ meV. Note however, that in the ice case 
the average water-molecule dipole moment is 3.36 D, which is even larger than
in the {\it homodromic} hexamer (3.05 D). This substantial induced dipole
moment can be ascribed to pronounced
polarization effects due to the presence of neighbor water molecules within
the 3-dimensional ice network.  

\subsection{Protonated water dimer}
The protonated water dimer H$_5$O$_2^+$ plays an
important role in the kinetics of aqueous solutions as well as in atmospheric 
chemistry.\cite{Dolenc} This system is characterized by a relatively flat
potential energy surface: in the global minimum, H$_2$O-H$^+$-H$_2$O,
the midpoint H nucleus, which bridges the two water molecules, is centered
approximately 1.2 \AA\ from each of the O atoms; however, there is another
stationary point on the potential energy surface, {H$_2$O...H$^+$-H$_2$O}, 
whose energy is only about
0.4 kcal/mol (= 0.02 eV) higher than the global minimum, 
where the midpoint H nucleus
is slightly closer to one of the O atoms.\cite{Dolenc} 
The former, most stable, H$_2$O-H$^+$-H$_2$O configuration can be 
described as obtained by the formation of 2 symmetric 
covalent bonds ({\it Zundel cation}), while the latter, 
H$_2$O...H$^+$-H$_2$O, is characterized by one HB and one covalent bond. 

Although the two different configurations are almost isoenergetic, their
differential electron charge densities (see Figs. 9 and 10) and
$\Delta \rho (z)$ profiles (shown in Fig. 11, where also the dimer curve
is reported for comparison) look quite different: in fact, the curve of 
Fig. 11, relative to H$_2$O...H$^+$-H$_2$O, qualitatively resembles 
that of water
dimer, but with much larger charge fluctuations, while $\Delta \rho (z)$ for
H$_2$O-H$^+$-H$_2$O exhibits a single pronounced maximum (typical of
conventional covalent bonds) centered around the midpoint H nucleus.
H$_2$O...H$^+$-H$_2$O is of particular interest because it can be considered
as a {\it strong} HB, being characterized by a binding energy of about 
1.5 eV,\cite{Steiner} which is almost an order of magnitude larger than that
of the ordinary water-dimer HB. Note that also the O...H$^+$ distance
is much shorter, 1.31 \AA, than that of O...H in water dimer, 1.92 \AA.

\subsection{HB characterization}
The prototypical HB in water dimer and HBs in other systems,
corresponding to wide range of bond strengths, can also
be quantitatively characterized by a single geometric bonding parameter, 
CCP, that is found to be correlated to the Mayer bond order
(MBO) parameter.\cite{Mayer,Bridgeman}
The MBO concept represents a 
popular tool to obtain insight into the strength and nature of bonding 
from first-principles
calculations because takes all the contributions to the bond into account,
although it is not devoid of limitations (see, for instance, ref. 
\onlinecite{Szarek}).
We have explicitly computed MBO parameters for the systems
investigated in the present study using the ab initio CPMD 
package.\cite{CPMD}

The {\it covalent character} parameter, CCP, 
is based on the positions (and their changes upon bonding)
of the centres, WFCs, of the MLWFs generated from the Kohn-Sham orbitals
(see Method section) and can be considered as an extrapolation
of a concept already introduced in the seminal paper of Marzari and
Vanderbilt\cite{Marzari} for the characterization of the {\it ionic} bonding:
``... the shift of the Wannier center away from the bond center might
serve as a kind of measure of bond ionicity.'' 
For instance, given our pseudopotential approximation, for a single water 
molecule we explicitly consider 8 valence electrons, the 2 additional 
oxygen-core electrons being essentially inert from the point of view of 
the bonding properties. 
Due to spin degeneracy we need therefore to consider 4 doubly-occupied
MLWFs and their relative WFCs.
These WFCs are tetrahedrally oriented, and two of them describe
the lone-pair orbitals while the other two represent covalent-bond
orbitals (see Fig. 3).\cite{psil1999}   
We also point out
that, particularly in weak and moderate HBs, the MLWF spreads are  
hardly affected by bonding, which represents a justification to focus 
on WFCs positions only: in fact, in the case of the water dimer, the spread
of the MLWF involved in the HB is changed by only 0.3\% from the value
of the corresponding, lone-pair MLWF of the isolated water molecule, and, 
even in hexagonal ice, the variation is smaller than 3\%.

By considering, as the reference system, always the water dimer investigated
above (the generalization to other H-bonded systems is easy) and referring
to the schematic, explanatory Fig. 3, in order to characterize the HB
between O$_a$ and H, we propose 
the following definition for the CCP parameter: 

\begin{equation}
{\rm CCP}=  (l-l_0)/(d_m-l_0) \;, 
\label{CCP}
\end{equation}
where $l$ and $l_0$ denote the distances of the W$_l$ WFC from O$_a$
in the H-bonded system and in the isolated water molecule, respectively,
and $d_m = d_b/2$ is half the HB distance between O$_a$ and H. 

Note that, on the basis of the density-differential analysis reported above, 
we expect that W$_l$ is pulled out due to the formation of the HB, 
so that $l > l_0$ (the same behavior is also observed in 
liquid water\cite{psil1999}).
Basically, according to this definition, CCP varies from 0 to 1: in the
limiting cases, CCP$=0$ (zero covalent contribution) if $l=l_0$, that is 
the distance of W$_l$ from O$_a$ is not changed upon bonding, while
CCP$=1$ (100\% covalent contribution) if $l=d_m$, so that W$_l$, 
which indicates the 
center-of-mass position of the interfragment electron charge, is precisely 
located at the geometric (midpoint) bond center, as, for instance, in the
H dimer, H$_2$.
Therefore our CCP represents indeed a simple estimate of 
the HB covalent character,
although it should be pointed out that a unique and commonly accepted 
definition of covalent contribution in a HB is still missing, so that
the HB covalency of water dimer has been given different weights, depending
on the different theoretical frameworks and definitions adopted   
(see, for instance ref. \onlinecite{Sterpone} and references therein).

In Table I we report the CCP values for different
systems (including those discussed above) covering a wide range 
of bond strengths, together with the corresponding MBO parameter, 
the HB energy, $E_b$, and bond length, $d_b$.
As far as the water dimer is concerned, at the equilibrium distance CCP=0.02,
so that the covalent contribution (2.0 \%) turns out to be rather small;
it is much smaller (essentially negligible) in the less favored,
{\it cyclic} and {\it bifurcated} water-dimer conformations, and when
the intermolecular distance is longer (O$_a$-O$_d$ separation=4.0 \AA)
than the equilibrium one. Interestingly, the 4.0 \AA\ distance has 
been indicated in the literature as the cutoff distance when any 
covalent contribution vanishes.\cite{Barbiellini,Sterpone}
In line with the differential-density analysis reported above, even 
at intermolecular distance shorter (O$_a$-O$_d$ separation=2.5 \AA) than
the equilibrium one, the covalent contribution is not much increased
(2.5 \%). Even for the HB in the methane-water compound (CH$_4$-H$_2$O ) 
covalency appears to be marginal.
As expected, for the cyclic water hexamer in the {\it homodromic} configuration,
the CCP value (averaged over the 6 HBs) is much larger than the
corresponding values both for the {\it antidromic} structure and for 
the water dimer,
thus showing that in this case the covalent contribution is much more
pronounced, again in agreement with the behavior of the differential-density
profiles and with the cooperative effects discussed above.  
CCP for ice in the hexagonal phase is comparable to that
of the cyclic water hexamer in the {\it homodromic} configuration.
The CCP values for the methanol dimer (CH$_3$OH-CH$_3$OH) and for the 
methanol hexamer ((CH$_3$OH)$_6$) in the
{\it ring} structure are very close to those for the water dimer and
water hexamer, respectively, in line with the behavior of the corresponding
HB binding energies. Note that methanol rings have been 
proposed\cite{Kashtanov} as good examples to illustrate the covalent 
contribution to the HB. 

Finally, the relatively high CCP values for H$_2$O...H$^+$-H$_2$O 
and FH-F$^-$, which represent {\it strong}
H-bonded systems (their HB energy is almost an order of magnitude 
larger that that of the water dimer), lead us to conclude that these complexes 
certainly possess substantial covalent 
character; in fact, they are sometimes  considered as a chemically 
bound species.\cite{Buckingham,Weinhold} 
Note that, for the Zundel cation H$_2$O-H$^+$-H$_2$O, CCP=0.357, thus
confirming the pronounced covalency of this system, as also discussed
in the previous subsection.

Data in Table I clearly show that there is a high correlation 
among the listed quantities. In particular
the {\it correlation coefficient} between $E_b$ and $d_b$ is
0.68, between $E_b$ and MBO is -0.89, between $E_b$ and CCP is -0.94,
and between MBO and CCP is 0.94, thus pointing out that, particularly
MBO and CCP are strictly correlated, but CCP better correlates with
$E_b$ than MBO. Hence
this simple geometric parameter, based on variations
of the WFCs positions, can be used to characterize a wide range of 
bond strengths, where
electrostatic, covalent, and dispersion contributions largely vary in their 
relative weights.

In Table II we report the values of MBO and 
CCP for the H$_2$O...H$^+$-H$_2$O system at different HB lengths, 
starting from the equilibrium O-O $d_{\rm OO}$ distance to 
larger distances (the geometry of the system is relaxed at fixed 
$d_{\rm OO}$ distance).
In this case the correlation coefficient between $E_b$ and $d_b$ is
0.94, between $E_b$ and MBO is -0.99, between $E_b$ and CCP is -0.84,
and between MBO and CCP is 0.89, thus confirming again the high
correlation between MBO and CCP even away from equilibrium.

\section{Conclusions}
The prototypical HB in water dimer and HBs in the protonated 
water dimer, in other small molecules, in water cyclic clusters, and in ice,
covering a wide range of bond strengths have
been investigated in terms of different structural, energetic, and 
electrostatic (induced molecular dipole moments) properties and
particularly focusing on the differential electron density and on the
shifts of the centres of the MLWFs.
The information from the
latter quantities can be conveyed into a single geometric bonding parameter,
CCP, that appears to be correlated to the MBO
and the HB strength, and gives an estimate of the HB covalent character.
For HBs in water (dimer, hexamers, ice) and methanol 
CCP ranges from 0.00 to 0.11 (MBO from 0.01 to 0.23), 
thus leading to the conclusion that, for such
moderate HBs, the covalent contribution is present but not predominant.
Test calculations on water dimer show that, at least for moderate-strength
HBs, using the standard PBE functional is appropriate since
adopting alternative DFT functionals that include vdW effects or are hybrid
in character only leads to small effects.
By considering the cyclic water hexamer and the hexagonal
phase of ice we have also elucidated the importance of 
cooperative/anticooperative effects in H-bonding formation. 


\vfill
\eject

\begin{table}
\caption{HB energy, $E_b$, bond length, $d_b$, MBO and CCP parameters,
for different H-bonded systems, listed in order of increasing MBO values. 
H$_2$O-H$_2$O-2.5\AA\ and H$_2$O-H$_2$O-4.0\AA\ indicate the water dimer
at O-O distances shorter and longer than the equilibrium one, 
H$_2$O-H$_2$O-{\it cyc} and H$_2$O-H$_2$O-{\it bif} the {\it cyclic} and 
{\it bifurcated}, less favored water-dimer conformations, 
(H$_2$O)$_6$-{\it homo} and (H$_2$O)$_6$-{\it anti} the
water hexamer in the {\it homodromic} and {\it antidromic} structures, and 
H$_2$O-{\it ice-hex} ice in the hexagonal phase.}
\begin{center}
\begin{tabular}{|l|r|c|c|c|}
\hline
               system   &$E_b$(meV)& $d_b$(\AA)& MBO & CCP \\ \tableline
\hline
 H$_2$O-H$_2$O-4.0\AA\  &  -89    &    3.02    &0.010&0.003 \\   
 H$_2$O-H$_2$O-{\it cyc}&  -81    &    2.33    &0.019&0.006 \\   
 CH$_4$-H$_2$O          &  -25    &    2.59    &0.021&0.003 \\
 H$_2$O-H$_2$O-{\it bif}&  -68    &    2.54    &0.022&0.004 \\   
 H$_2$O-H$_2$O          & -221    &    1.92    &0.105&0.020 \\   
 (H$_2$O)$_6$-{\it anti}& -228    &    1.92    &0.124&0.040 \\
 CH$_3$OH-CH$_3$OH      & -232    &    1.89    &0.124&0.041 \\
 H$_2$O-H$_2$O-2.5\AA\  &  -74    &    1.54    &0.182&0.025 \\   
 H$_2$O-{\it ice-hex}   & -328    &    1.75    &0.191&0.083 \\
 (H$_2$O)$_6$-{\it homo}& -381    &    1.62    &0.222&0.105 \\
 (CH$_3$OH)$_6$         & -389    &    1.59    &0.226&0.105 \\
 H$_2$O...H$^+$-H$_2$O  &-1651    &    1.31    &0.366&0.172 \\
 FH-F$^-$               &-1969    &    1.16    &0.421&0.290 \\
\hline
\end{tabular}
\end{center}
\label{tableh-bond}
\end{table}
\vfill
\eject

\begin{table}
\caption{O-O distance, $d_{\rm OO}$, bond length, $d_b$, 
HB energy, $E_b$, MBO and CCP parameters for the 
H$_2$O...H$^+$-H$_2$O system, at different $d_{\rm OO}$ distances.}
\begin{center}
\begin{tabular}{|c|c||r|c|c|}
\hline
$d_{\rm OO}$ (\AA)& $d_b$(\AA)& $E_b$(meV)&  MBO  & CCP \\ \tableline
\hline
   2.38           & 1.31      &  -1651    & 0.366 &0.172\\
   2.65           & 1.55      &  -1458    & 0.298 &0.056\\
   2.91           & 1.87      &  -1154    & 0.196 &0.044\\
   3.18           & 2.16      &   -903    & 0.126 &0.027\\
   3.44           & 2.44      &   -714    & 0.078 &0.016\\
   3.70           & 2.71      &   -573    & 0.046 &0.015\\
   3.97           & 2.98      &   -468    & 0.026 &0.011\\
   4.23           & 3.24      &   -389    & 0.014 &0.002\\
   5.29           & 4.31      &   -208    & 0.001 &0.003\\
\hline
\end{tabular}
\end{center}
\label{tabledistances}
\end{table}
\vfill
\eject

\begin{figure}
\centerline{
\includegraphics[width=15cm,angle=180]{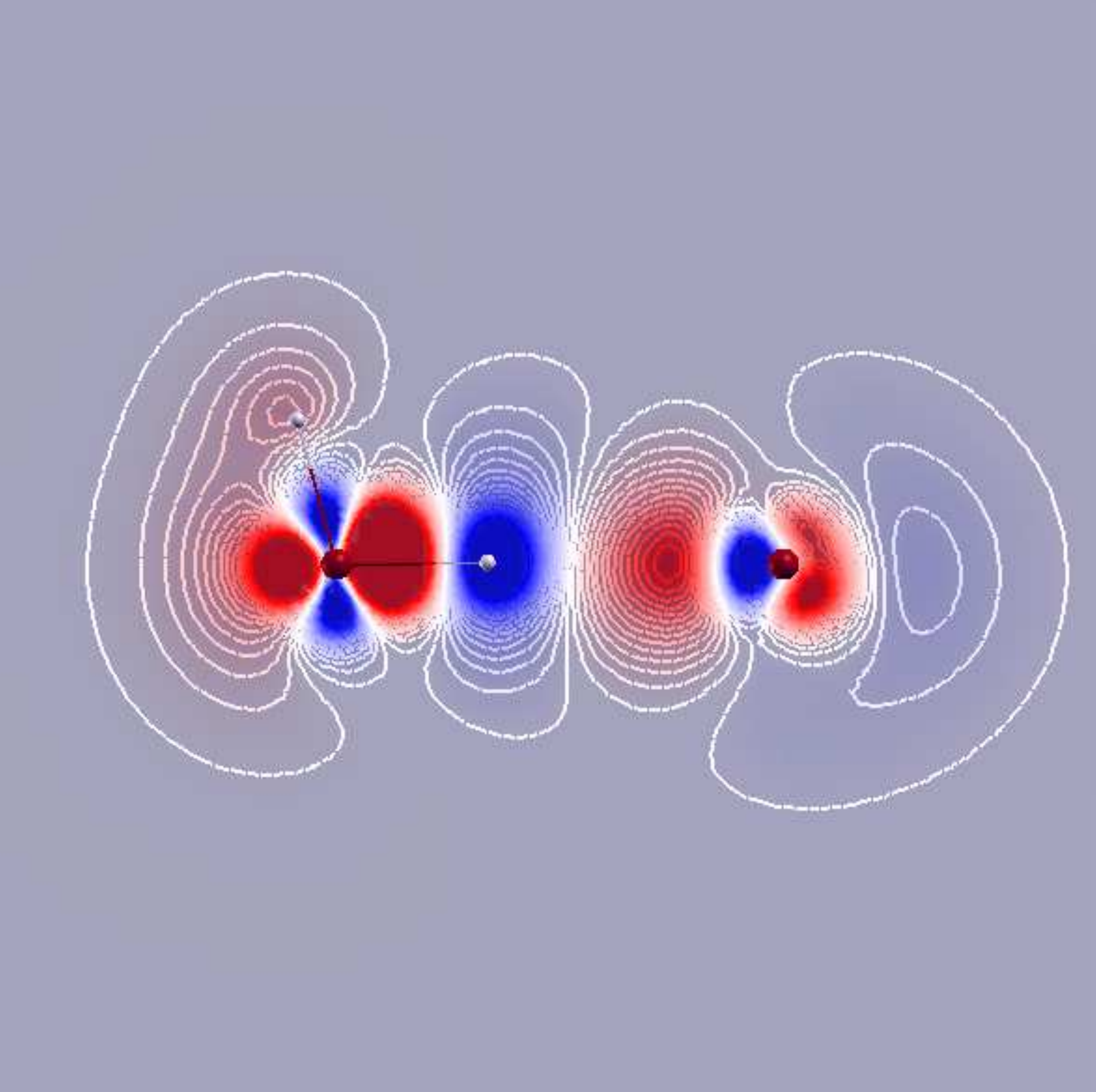}
}
\caption{Differential electron charge density, $\Delta \rho$,
 for the water dimer in the optimal, linear HB configuration 
(shown is the difference
between the dimer density and those of the isolated water molecules),
plotted on a plane containing the HB.
The red and white balls denote O and H atoms, respectively.
Red areas indicate electron density gain, while blue areas indicate loss
of electron density; increments between contour lines correspond to
$2\times 10^{-4}\, e/{\rm \AA}^3$ 
(within the range $\pm 0.003\, e/{\rm \AA}^3$).}
\label{figdimer2D}
\end{figure}
\eject
                      
\begin{figure}
\centerline{
\includegraphics[width=15cm,angle=-90]{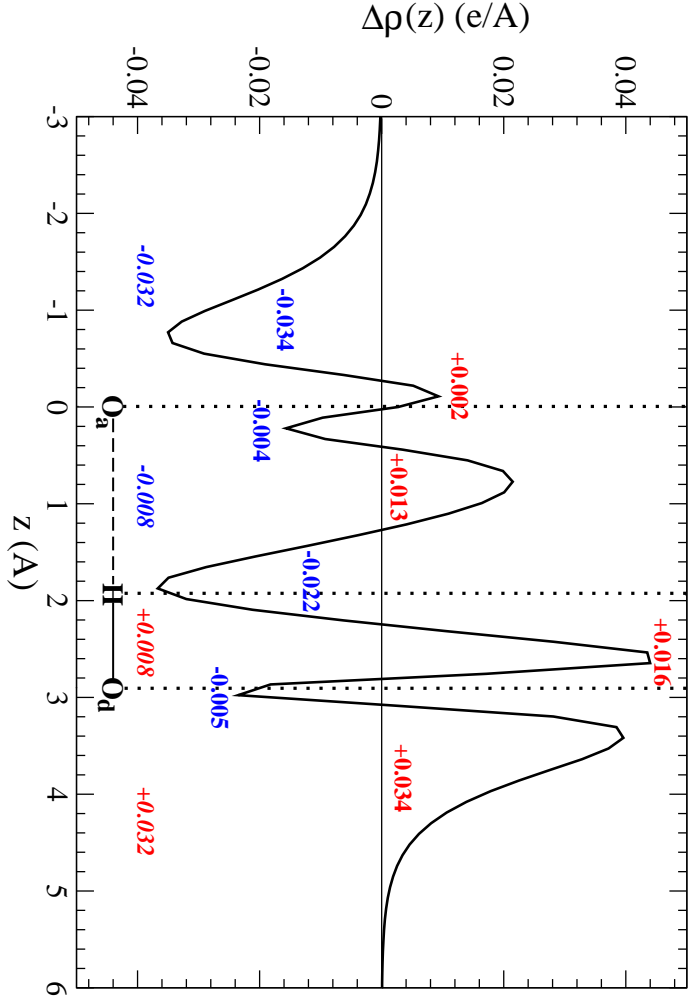}
}
\caption{Differential electron charge density, along the $z$ axis, 
$\Delta \rho (z)$, integrated over $x,y$ planes (see text for the definition) 
for the water dimer in the optimal, linear HB configuration 
(corresponding to Fig. 1)
Positive (in red) and negative (in blue) numerical values inside 
or near the peaks indicate 
the corresponding electron charge (in $e$) obtained by integral of
$\Delta \rho (z)$ along $z$; the numerical values in italics
in the bottom indicate the electron charge below the O atom of the
{\it acceptor} molecule O$_a$, between O$_a$ and H (the HB region),
between H and O$_d$ (the covalent bond region), and above the O atom of the
{\it donor} molecule O$_d$, respectively (these 4 regions are delimited 
by vertical, dotted lines).}   
\label{figdimer1D}
\huge
\end{figure}
\eject

\begin{figure}
\centerline{
\includegraphics[width=15cm]{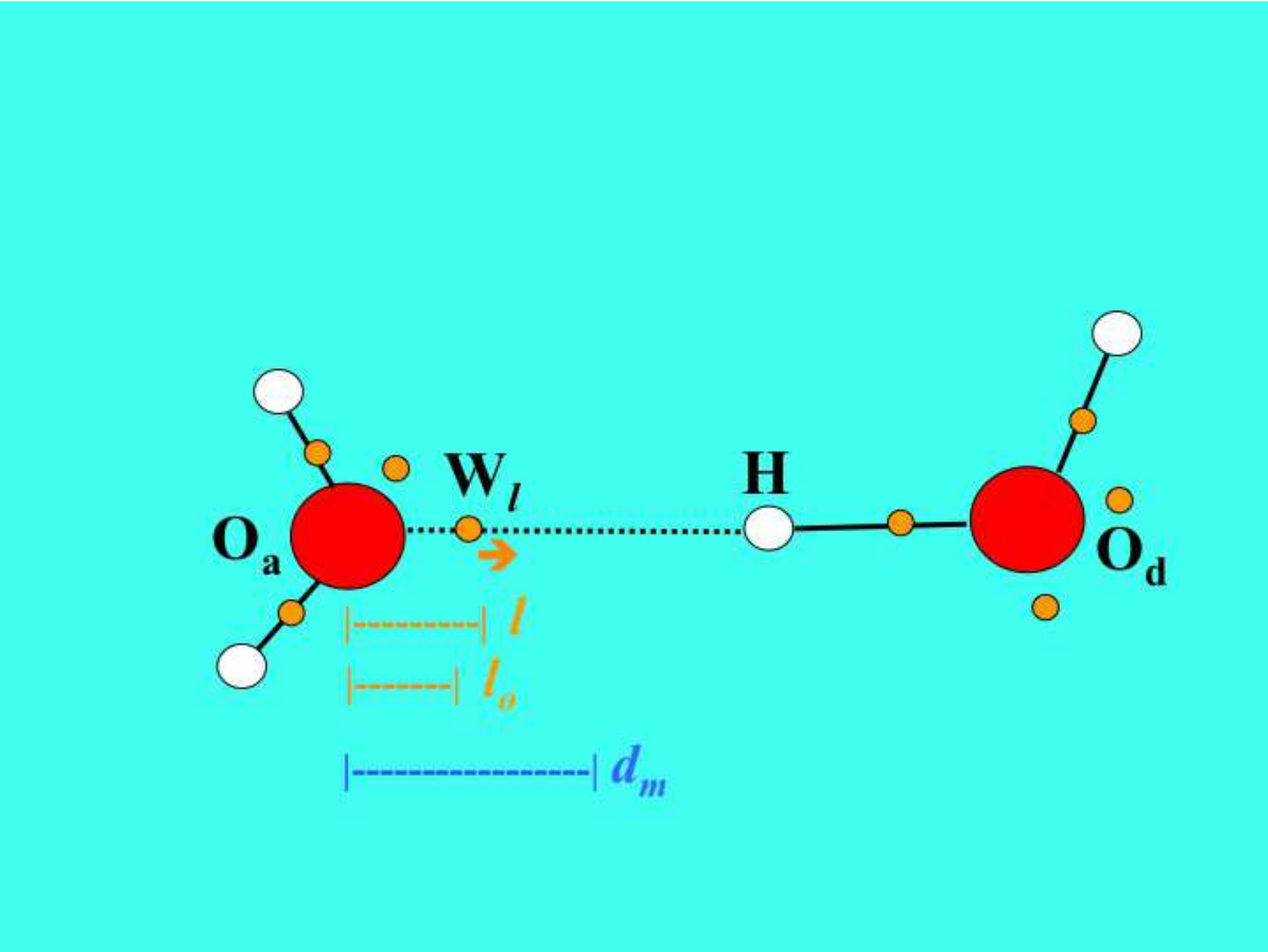}
}
\caption{Explanatory figure with schematic distributions of 
atoms (red balls: O atoms of the donor, $O_d$, and acceptor, $O_a$, molecule;
white balls: H atoms) and WFCs (orange balls) involved in the definitions of
the CCP parameter (see text), for the water dimer.} 
\label{figCC}
\huge
\end{figure}
\eject

\begin{figure}
\centerline{
\includegraphics[width=15cm,angle=-90]{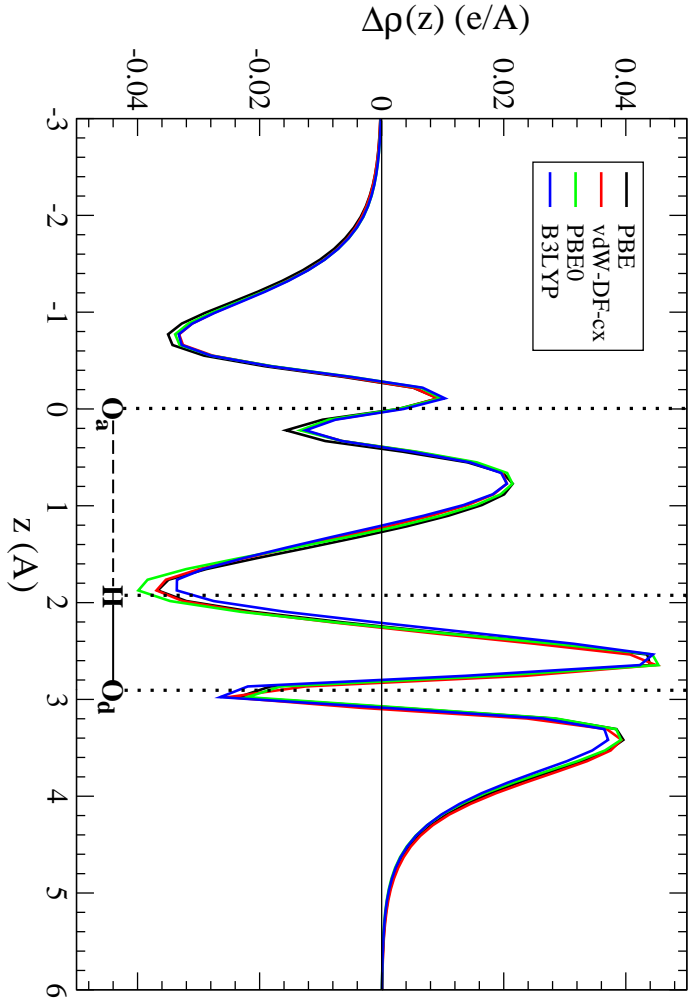}
}
\caption{Differential electron charge density, along the $z$ axis, 
$\Delta \rho (z)$, integrated over $x,y$ planes (see Fig. 2) obtained
using different DFT functionals: PBE (GGA), vdW-DF-cx (vdW-corrected DFT),
and two hybrid functionals, PBE0 and B3LYP.} 
\label{figdimer1Dfunctionals}
\huge
\end{figure}
\eject

\begin{figure}
\centerline{
\includegraphics[width=15cm,angle=-90]{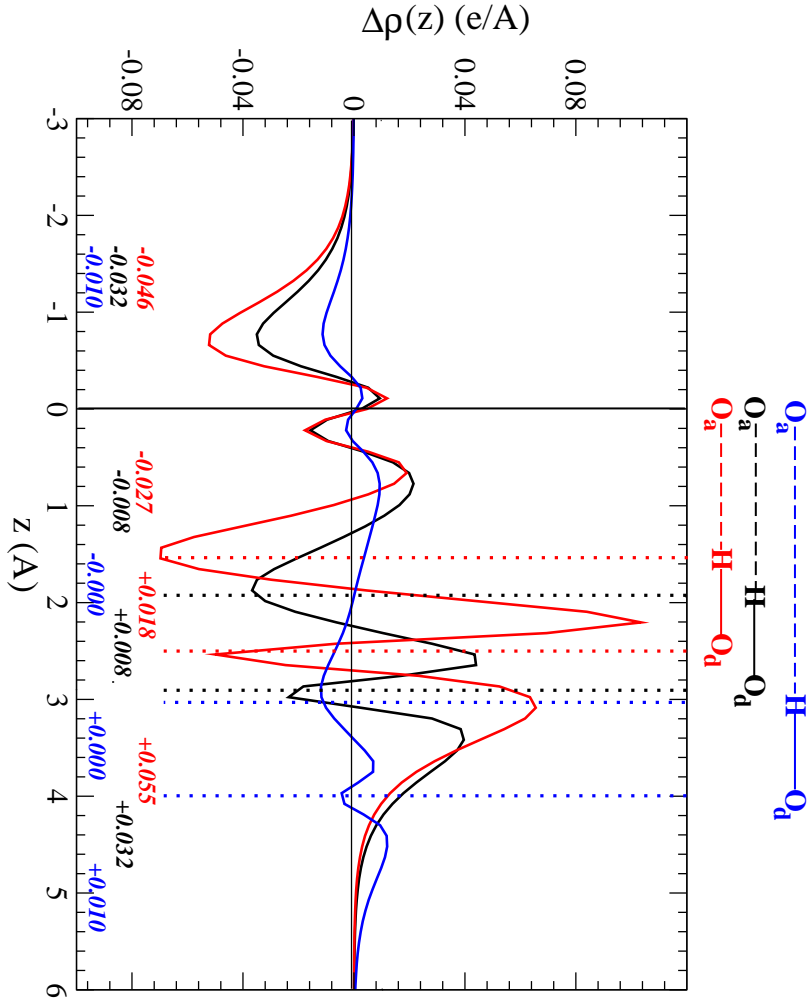}
}
\caption{Differential electron charge density, along the $z$ axis, 
$\Delta \rho (z)$, integrated over $x,y$ planes (see Fig. 2) obtained
at PBE level and considering a water-water distance shorter 
(O-O distance = 2.5 \AA, red curve) or longer (O-O distance = 4.0 \AA, blue
curve) than the equilibrium value (O-O distance = 2.91 \AA, black curve).
Positive and negative numerical values indicate 
the electron charge (in $e$) below the O atom of the
{\it acceptor} molecule O$_a$, between O$_a$ and H (the HB region),
between H and O$_d$ (the covalent bond region), and above the O atom of the
{\it donor} molecule O$_d$, respectively (these 4 regions are delimited 
by vertical, dotted lines).}   
\label{figdimer1Ddistances}
\huge
\end{figure}
\eject

\begin{figure}
\centerline{
\includegraphics[width=15cm]{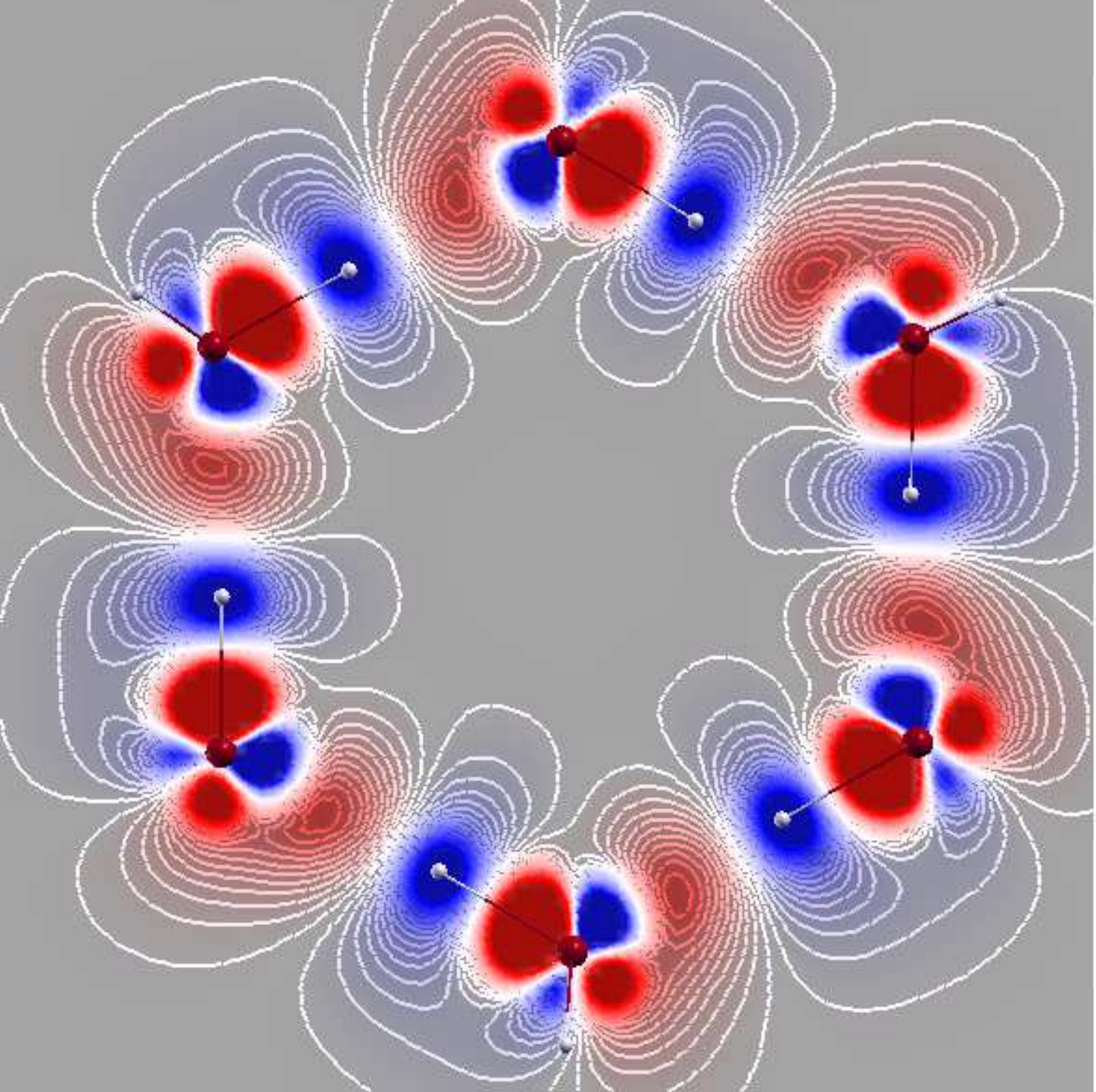}
}
\caption{Differential electron charge density, $\Delta \rho$,
for the water hexamer
in the cyclic {\it homodromic} configuration (shown is the difference
between the dimer density and those of the isolated water molecules),
plotted on a plane containing the hexamer ring and the HBs.
The red and white balls denote O and H atoms, respectively.
Red areas indicate electron density gain, while blue areas indicate loss
of electron density; increments between contour lines correspond to
$7\times 10^{-4}\, e/{\rm \AA}^3$ 
(within the range $\pm 0.01\, e/{\rm \AA}^3$).}
\label{fighexamer2D}
\end{figure}
\eject
                      
\begin{figure}
\centerline{
\includegraphics[width=15cm]{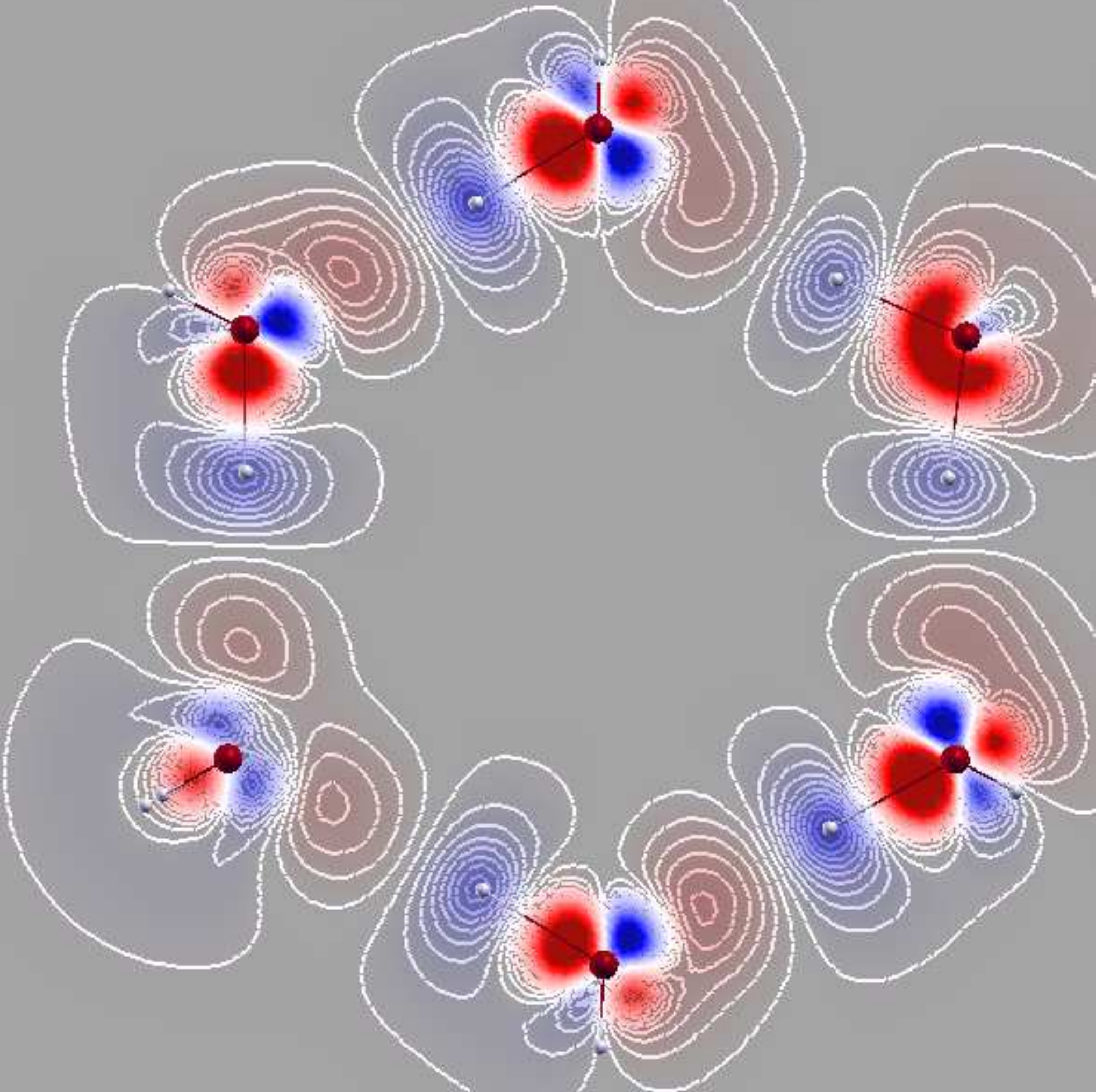}
}
\caption{Differential electron charge density, $\Delta \rho$,
for the water hexamer
in the cyclic {\it antidromic} configuration (shown is the difference
between the dimer density and those of the isolated water molecules),
plotted on a plane containing the hexamer ring and the HBs.
The red and white balls denote O and H atoms, respectively.
Red areas indicate electron density gain, while blue areas indicate loss
of electron density; increments between contour lines correspond to
$7\times 10^{-4}\, e/{\rm \AA}^3$ 
(within the range $\pm 0.01\, e/{\rm \AA}^3$).}
\label{fighexameranti2D}
\end{figure}
\eject
                      
\begin{figure}
\centerline{
\includegraphics[width=15cm,angle=-90]{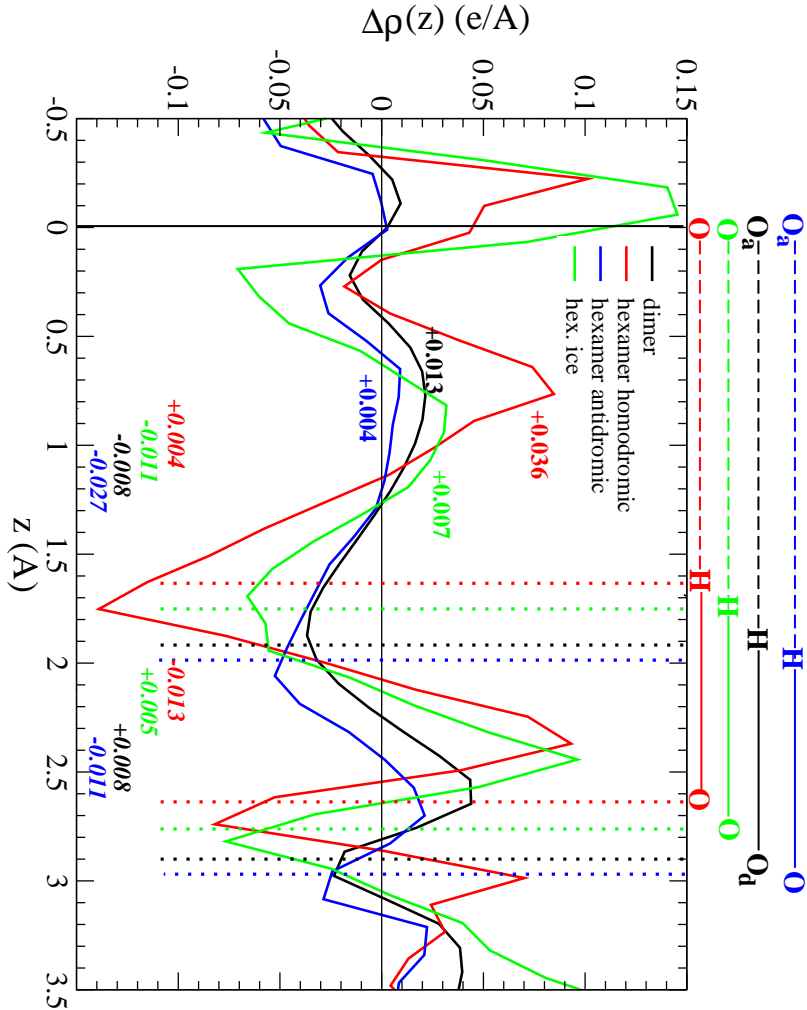}
}
\caption{Differential electron charge density, $\Delta \rho (z)$, 
for the water dimer, 
2 cyclic water hexamers ({\it homodromic} and {\it antidromic} case), 
and hexagonal ice.
Positive numerical values near the second peak indicate accumulation of
the corresponding electron charge (in $e$) obtained by integral of
$\Delta \rho (z)$ along $z$; the numerical values in italics
in the bottom indicate 
the electron charge (in $e$) between O and H (the HB region) and
between H and O (the covalent bond region), respectively 
(these regions are delimited by vertical, dotted lines).} 
\label{fighexamer1D}
\end{figure}
\eject

\begin{figure}
\centerline{
\includegraphics[width=15cm]{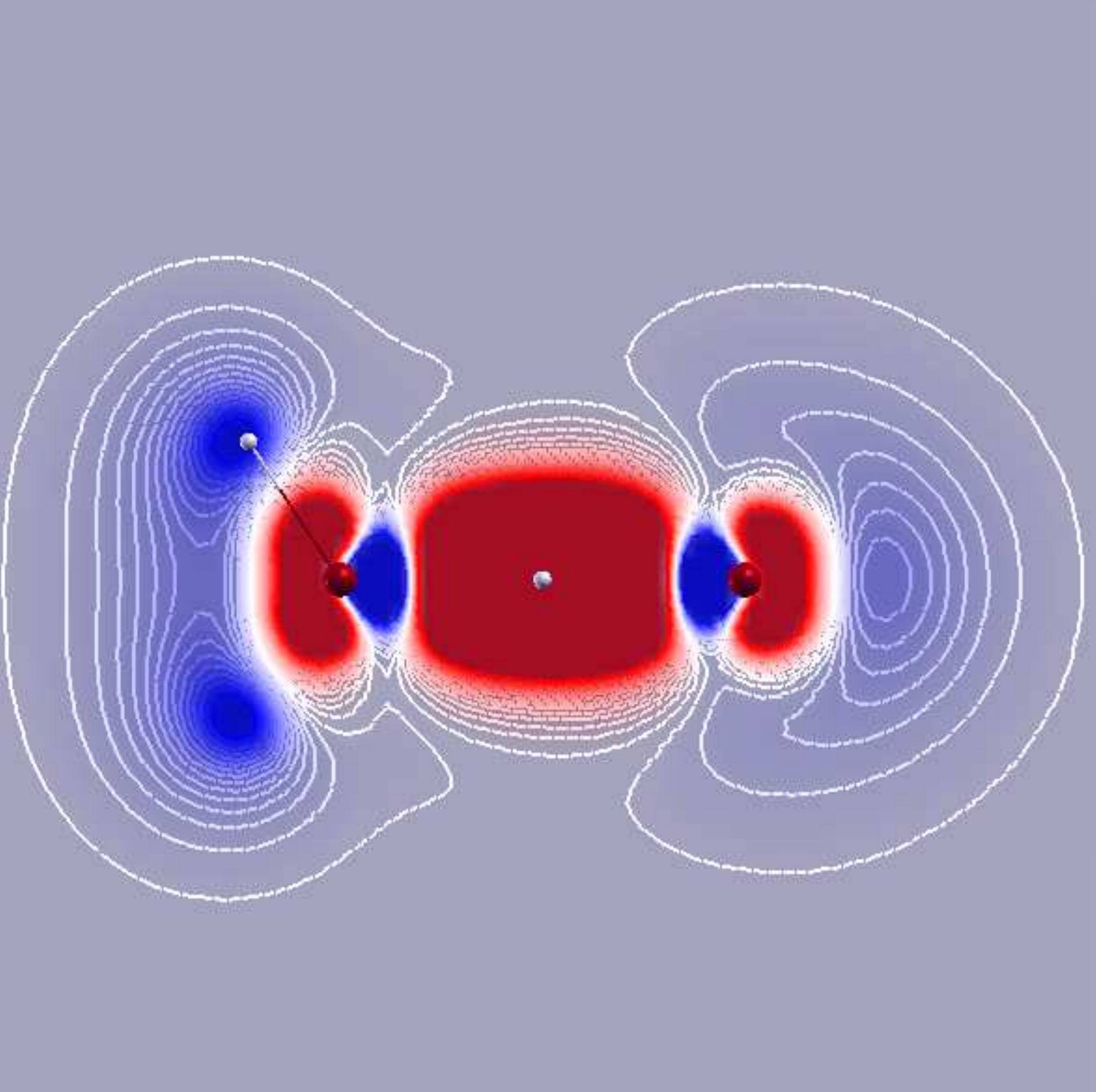}
}
\caption{Differential electron charge density, $\Delta \rho$, 
for the protonated water dimer
in the H$_2$O-H$^+$-H$_2$O (Zundel cation) configuration (shown is 
the difference between the dimer density and those of the isolated 
water molecules), plotted on a plane containing the two O atoms and H$^+$.
Red areas indicate electron density gain, while blue areas indicate loss
of electron density; increments between contour lines correspond to
$7\times 10^{-4}\, e/{\rm \AA}^3$ 
(within the range $\pm 0.01\, e/{\rm \AA}^3$).}
\label{figzundel2D}
\end{figure}
\eject
                      
\begin{figure}
\centerline{
\includegraphics[width=15cm]{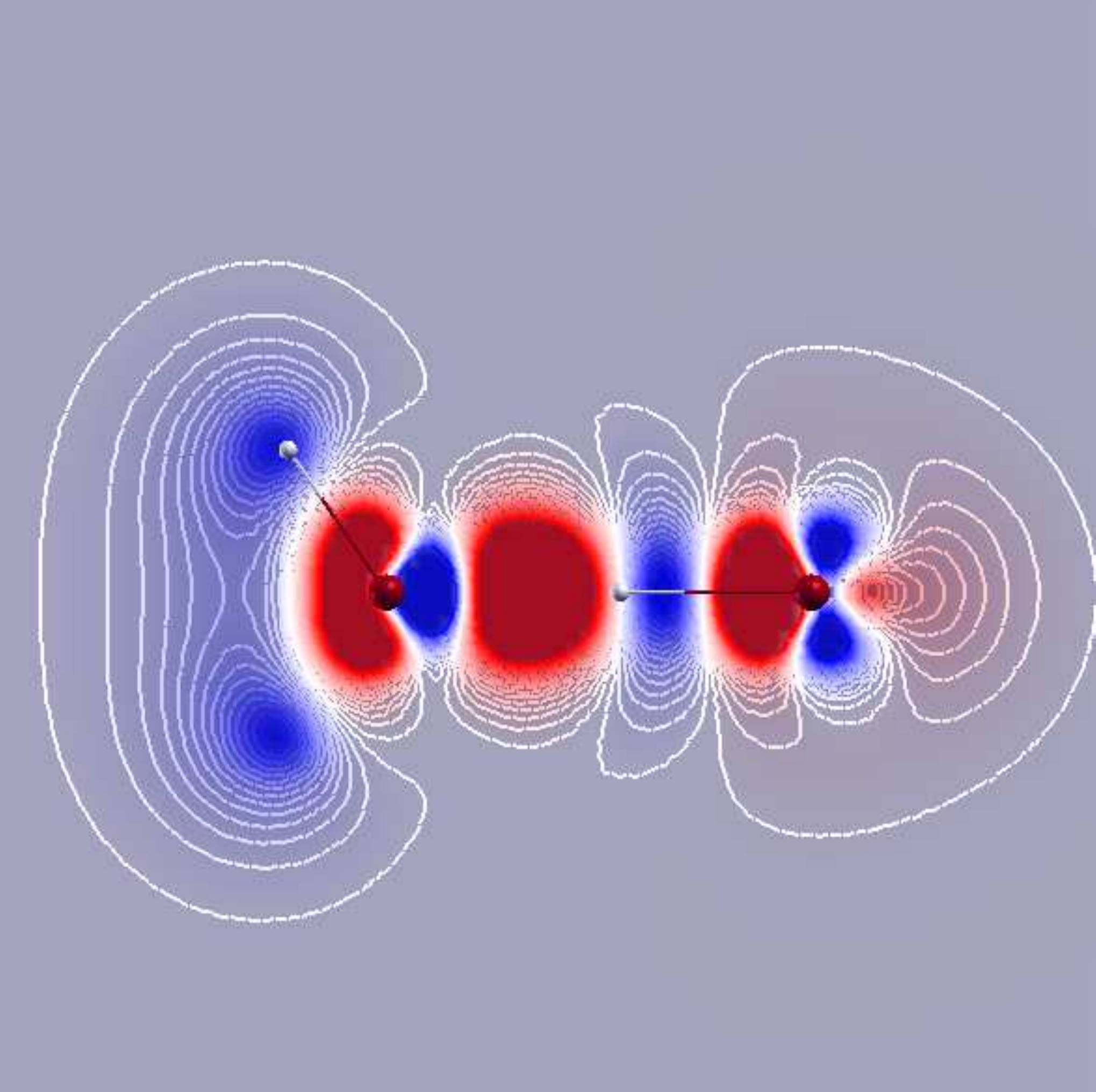}
}
\caption{Differential electron charge density, $\Delta \rho$, 
for the protonated water dimer
in the H$_2$O...H$^+$-H$_2$O configuration (shown is 
the difference between the dimer density and those of the isolated 
water molecules), plotted on a plane containing the two O atoms and H$^+$.
Red areas indicate electron density gain, while blue areas indicate loss
of electron density; increments between contour lines correspond to
$7\times 10^{-4}\, e/{\rm \AA}^3$ 
(within the range $\pm 0.01\, e/{\rm \AA}^3$).}
\label{figprotwat2D}
\end{figure}
\eject
                      
\begin{figure}
\centerline{
\includegraphics[width=15cm,angle=-90]{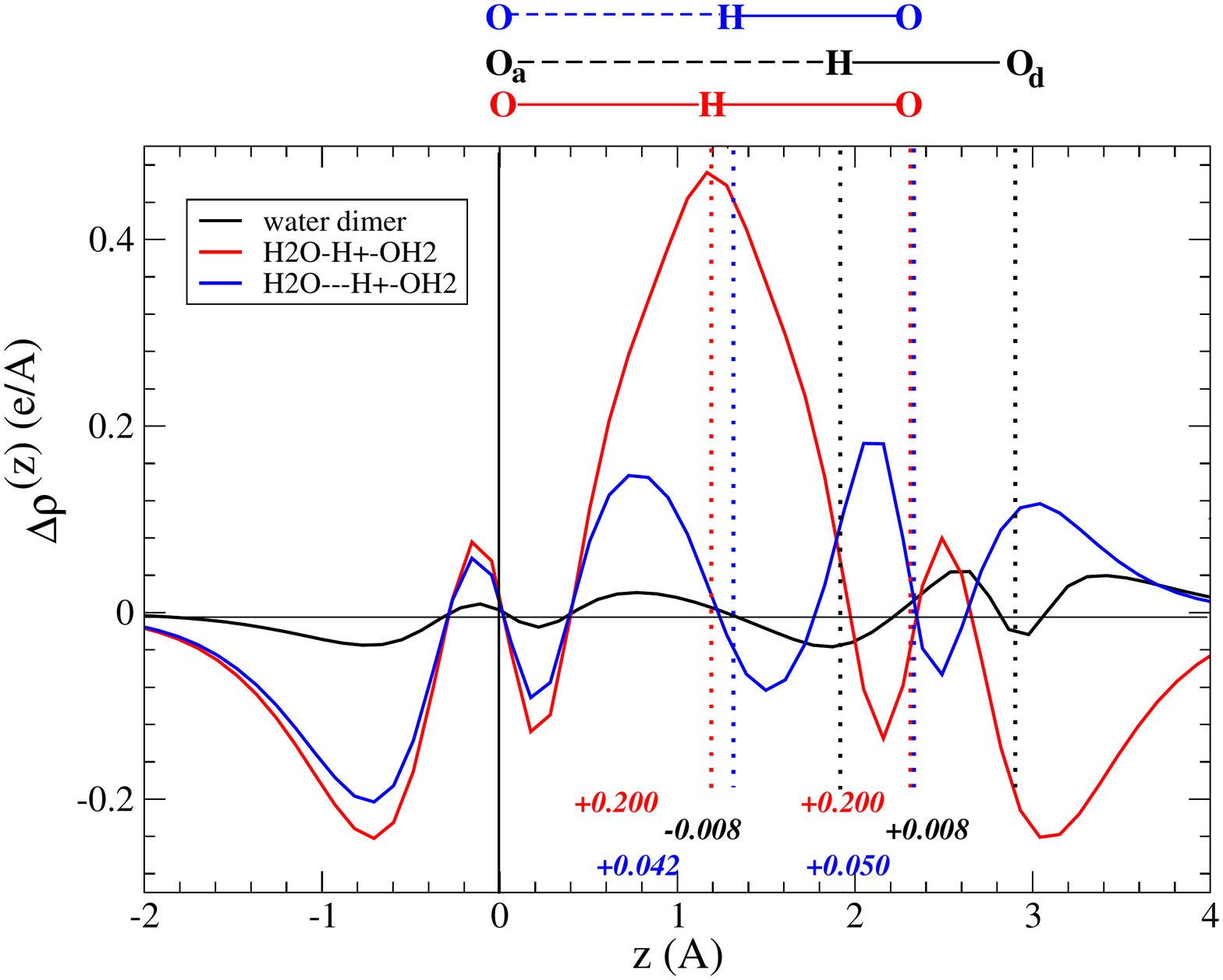}
}
\caption{Differential electron charge density, $\Delta \rho (z)$, 
for the water dimer and the
protonated water dimer in two different configurations, H$_2$O-H$^+$-H$_2$O
(the Zundel cation) and H$_2$O...H$^+$-H$_2$O.
The numerical values in the bottom indicate 
the electron charge (in $e$) between O and H, and between H and O, that
are the HB and the covalent bond region, respectively, for the water dimer
and H$_2$O...H$^+$-H$_2$O
(these regions are delimited by vertical, dotted lines).} 
\label{figH5O2+1D}
\end{figure}
\eject


\begin{thebibliography}{9}
\bibitem{Jeffrey} G. A. Jeffrey, {\it An Introduction to
                  Hydrogen bonding}, Oxford University Press,
                  New York (1997).
\bibitem{Walrafen} G. E. Walrafen, M. R. Fisher, M. S. Hokmabadi, W.-H. Yang,
                   J. Chem. Phys. {\bf 85}, 6970 (1986).
\bibitem{Buckingham} A. D. Buckingham, J. E. Del Bene, S. A. C. McDowell,
                     Chem. Phys. Lett. {\bf 463}, 1 (2008).
\bibitem{Steiner} T. Steiner, 
                  Angew. Chem. Int. Ed. {\bf 41}, 48 (2002).
\bibitem{BWang} B. Wang, M. Xin, X. Dai, R. Song, Y. Meng, J. Han, 
                W. Jiang, Z. Wang, R. Zhang,
                Phys. Chem. Chem. Phys {\bf 17}, 2987 (2015);
                B. Wang, L. Wang, X. Dai, Y. Gao, W. Jiang, J. Han, 
                Z. Wang, R.-Q. Zhang,
                Int. J. Quantum Chem. {\bf 115}, 817 (2015).
\bibitem{Barbiellini} E. Isaacs, A. Shukla, P. Platzman, D. Hamann,
                      B. Barbiellini, C. Tulk, 
                      Phys. Rev. Lett. {\bf 82}, 600 (1999);
                      B. Barbiellini, A. Shukla,
                      Phys. Rev. B {\bf 66}, 235101 (2002).
\bibitem{Romero} A. H. Romero, P. L. Silvestrelli, M. Parrinello,
                 J. Chem. Phys. {\bf 115}, 115 (2001).
\bibitem{Weinhold} F. Weinhold, R. A. Klein, 
                   Mol. Phys. {\bf 110}, 565 (2012).
\bibitem{Chaplin} M. F. Chaplin, {\it Water's hydrogen bond strength}, 
                  In: {\it Water and Life}, ed. R. M. Lynden-Bell, 
                  S. Conway Morris, J. D. Barrow, J. L. Finney and 
                  C. L. Harper, Jr. (CRC Press, Boca Raton, 2010) 
                  pp 69-86; arXiv:0706.1355 (2007). 
\bibitem{Sterpone} F. Sterpone, L. Spanu, L. Ferraro, S. Sorella,
                   L. Guidoni,
                   J. Chem. Theory Comput. {\bf 4}, 1428 (2008).
\bibitem{Elgabarty} H. Elgabarty, R. Z. Khaliullin, T. D. K\"uhne,
                    Nature Communications {\bf 6}, 8318 (2015).
\bibitem{BWangSR} B. Wang, W. Jiang, X. Dai, Y. Gao, Z. Wang, R.-Q. Zhang,
               Scientific Reports {\bf 6}, 22099 (2016).
\bibitem{Zhang} J. Zhang, P. Chen, B. Yuan, W. Ji, Z. Cheng, X. Qiu,
                Science {\bf 342}, 611 (2013).
\bibitem{NBO} A. E. Reed, F. Weinhold, L. A. Curtiss, D. J. Pochatko,
              J. Chem. Phys. {\bf 84}, 5687 (1986).
\bibitem{Stone} A. J. Stone,
                J. Phys. Chem. A {\bf 121}, 1531 (2017).
\bibitem{Khaliullin} R. Z. Khaliullin, E. A. Cobar, A. T. Bell, 
                     M. Head-Gordon,
                     J. Phys. Chem. A {\bf 111}, 8753 (2007);
                     R. Z. Khaliullin, A. T. Bell, M. Head-Gordon,
                     J. Chem. Phys. {\bf 128}, 184112 (2008).
\bibitem{Santra1} B. Santra, A. Michaelides, M. Scheffler,
                  J. Chem. Phys. {\bf 127}, 184104 (2007).
\bibitem{Santra2} B. Santra, A. Michaelides, M. Fuchs, A. Tkatchenko,
                  C. Filippi, M. Scheffler,
                  J. Chem. Phys. {\bf 129}, 194111 (2008).
\bibitem{BLYP} A. D. Becke,
               Phys. Rev. A {\bf 38}, 3098 (1988);
               C. Lee, W. Yang, R. C. Parr,
               Phys. Rev. B {\bf 37}, 785 (1988).
\bibitem{PBE} J. P. Perdew, K. Burke, M. Ernzerhof,
              Phys. Rev. Lett. {\bf 77}, 3865 (1996).
\bibitem{mycpl} P. L. Silvestrelli,
                Chem. Phys. Lett. {\bf 475}, 285 (2009). 
\bibitem{Arey} J. S. Arey, P. C. Aeberhard, I.-C. Lin, U. Rothlisberger,
               J. Phys. Chem. B {\bf 113}, 4726 (2009).
\bibitem{Kohn} See, for instance, W. Kohn, Y. Meir, D. E. Makarov,
               Phys. Rev. Lett. {\bf 80}, 4153 (1998).
\bibitem{HK} P. Hohenberg, W. Kohn,
             Phys. Rev. {\bf 136}, B864 (1964).
\bibitem{Lane} J. R. Lane, J. Contreras-Garc\'ia, J. P. Piquemal, B. J. Miller,
               H. G. Kjaergaard,
               J. Chem. Theory Comput. {\bf 9}, 3263 (2013).      
\bibitem{Marzari} N. Marzari, D. Vanderbilt,
                  Phys. Rev. B {\bf 56}, 12847 (1997).
\bibitem{psil1999} P. L. Silvestrelli, M. Parrinello,
                   Phys. Rev. Lett. {\bf 82}, 3308 (1999);
                   J. Chem. Phys. {\bf 111}, 3572 (1999).
\bibitem{Silvestrelli} P. L. Silvestrelli,
                       Phys. Rev. Lett. {\bf 100}, 053002 (2008);
                       J. Phys. Chem. A {\bf 113}, 5224 (2009).
\bibitem{ESPRESSO} P. Giannozzi, S. Baroni, N. Bonini, M. Calandra, R. Car, 
                   C. Cavazzoni, D. Ceresoli, G. L. Chiarotti, M. Cococcioni, 
                   I. Dabo, A. Dal Corso, S. Fabris, G. Fratesi, 
                   S. de Gironcoli, R. Gebauer, U. Gerstmann, C. Gougoussis, 
                   A. Kokalj, M. Lazzeri, L. Martin-Samos, N. Marzari, 
                   F. Mauri, R. Mazzarello, S. Paolini, A. Pasquarello,
                   L. Paulatto, C. Sbraccia, S. Scandolo, G. Sclauzero, 
                   A. P. Seitsonen, A. Smogunov, P. Umari, R. M. Wentzcovitch, 
                   J.Phys.: Condens. Matter {\bf 21}, 395502 (2009);
                   http://arxiv.org/abs/0906.2569. 
\bibitem{WanT} WanT code by A. Ferretti, B. Bonferroni, A. Calzolari, and 
               M. Buongiorno Nardelli, http://www.wannier-transport.org ;
               see also A. Calzolari, N. Marzari, I. Souza,
               M. Buongiorno Nardelli, Phys. Rev. B {\bf 69}, 035108 (2004).
\bibitem{Galvez} O. G\'alvez, P. C. G\'omez, L. F. Pacios,
                 J. Chem. Phys. {\bf 115}, 11166 (2001).
\bibitem{Lee} A. J. Lee, S. W. Rick,
              J. Chem. Phys. {\bf 134}, 184507 (2011).
\bibitem{Jiang} H. Jiang, O. A. Moultos, I. G. Economou, A. Z.
                Panagiotopoulos, 
                J. Phys. Chem. B {\bf 120}, 12538 (2016).
\bibitem{Nilsson} A. Nilsson, H. Ogasawara, M. Cavalleri, D. Nordlund,
                  M. Nyberg, Ph. Wernet, L. G. M. Pettersson,
                  J. Chem. Phys. {\bf 122}, 154505 (2005).
\bibitem{Tschumper} G. S. Tschumper, M. L. Leininger, B. C. Hoffman,
                    E. F. Valeev, H. F. Schaefer III, M. Quack,
                    J. Chem. Phys. {\bf 116}, 690 (2002).
\bibitem{BerlandPRB} K. Berland, P. Hyldgaard,
                     Phys. Rev. B {\bf 89}, 035412 (2014).
\bibitem{BerlandJCP} K. Berland, C. A. Arter, V. R. Cooper, 
                     K. Lee, B. I. Lundqvist, E. Schr\"oder, 
                     T. Thonhauser, P. Hyldgaard,
                     J. Chem. Phys. {\bf 140}, 18A539 (2014).
\bibitem{PBE0} C. Adamo, V. Barone,
               J. Chem. Phys. {\bf 110}, 6158 (1999).
\bibitem{B3LYP} A. D. Becke, 
                J. Chem. Phys. {\bf 98}, 5648 (1993);
                P. J. Stephens, F. J. Devlin, C. F. Chabalowski, 
                M. J. Frisch,
                J. Phys. Chem. {\bf 98}, 11623 (1994).
\bibitem{Fuster} F. Fuster, B. Silvi, 
                 Theor. Chem. Acc. {\bf 104}, 13 (2000).
\bibitem{Katsyuba} S. A. Katsyuba, M. V. Vener, E. E. Zvereva, 
                   J. G. Brandenburg,
                   Chem. Phys. Lett. {\bf 672}, 124 (2017).
\bibitem{Perez} C. P\'erez, D. P. Zaleski, N. A. Seifert, B. Temelso,
                G. C. Shields, Z. Kisiel, B. H. Pate,
                Angew. Chem. Int. Ed. {\bf 53}, 14368 (2014).  
\bibitem{Ohno} K. Ohno, M. Okimura, N. Akai, Y. Katsumoto, 
               Phys. Chem. Chem. Phys. {\bf 7}, 3005 (2005).
\bibitem{Saenger} W. Saenger,
                  Nature {\bf 279}, 343 (1979).
\bibitem{Kashtanov} S. Kashtanov, A. Augustson, J.-E. Rubensson, 
                    J. Nordgren, H. \AA gren, J.-H. Guo, Y. Luo, 
                    Phys. Rev. B {\bf 71}, 104205 (2005).
\bibitem{Dolenc} J. Dolenc, J. Koller,
                 Acta Chim. Slov. {\bf 53}, 229 (2006).
\bibitem{Mayer} I. Mayer, 
                Chem. Phys. Lett. {\bf 97}, 270 (1983).
\bibitem{Bridgeman} A. J. Bridgeman, G. Cavigliasso, L. R. Ireland,
                    J. Rothery,
                    J. Chem. Soc., Dalton Trans. 2095 (2001). 
\bibitem{Szarek} P. Szarek, Y. Sueda, A. Tachibana,
                 J. Chem. Phys. {\bf 129}, 094102 (2008).
\bibitem{CPMD} http://www.cpmd.org/
\end{thebibliography}
\end{document}